\documentclass[prc,showpacs,floatfix,twocolumn,superscriptaddress,nofootinbib]
{revtex4}
\usepackage{colordvi}

\usepackage{graphics}
\usepackage{longtable}
\usepackage{epsfig}
\usepackage{amsmath,amssymb,amsfonts}

\begin{document}

\title{The ratios between the moments of inertia of triaxial nuclei: Comment on the Physical Review C  95, 064315 (2017) }
\vskip 10pt

\author{S. Frauendorf}
\affiliation{Department of Physics, University of Notre Dame, South Bend, Indiana 46556, USA}

\begin{abstract}
{}

\end{abstract}

\maketitle

In Ref. \cite{Tanabe17},  K. Tanabe and K. Sugawara-Tanabe claim that the transverse wobbling mode, suggested in Ref. \cite{FD14},
does not exist. In Refs. \cite{Tanabe10,Tanabe17} they suggest an alternative model to explain the observed rotational bands. This comment
exposes my concerns about their work.

The appearance of wobbling excitations has been suggested as  a hallmark for    quantal rotation of triaxial nuclei \cite{Bo75}.
Experimental evidence for wobbling in the presence of an odd i$_{13/2}$ proton has been found in $^{163}$Lu \cite{Od01} and
in the presence of an odd h$_{11/2}$ proton in $^{135}$Pr \cite{Matta15}. The observations have been interpreted by coupling
 the odd proton with a triaxial rotor that describes the even-even core
 \cite{Ha02,Tanabe10,Tanabe17,FD14,Matta15}. The description sensitively depends on the ratios between the three moments of inertia of the triaxial rotor.
 These ratios are restricted by the indistinguishability of the protons an neutrons which constitute the rotor core. 
 Quantal rotation about a symmetry axis is not possible \cite{Bo75}, i. e. the moment of inertia of a symmetry axis is zero. More generally,
 the larger the deviation  of the density distribution from symmetry with respect to one of its principal axes the larger  the  moment of inertia.
 This implies that the medium axis has the largest moment of inertia. This fundamental property of the quantal many body system is in stark contrast
 with the classical rigid body  values of the moments of inertia of the triaxial density distribution
 \begin{equation}
{ \cal J}^{rig}_k={\cal J}_0\left[1-\beta\left(\frac{4}{4\pi}\right)^{1/2}\cos\left(\gamma+\frac{2}{3}\pi k\right)\right],
  \end{equation}
  while the irrotational flow values of an ideal liquid,
 \begin{equation}
{ \cal J}^{hyd}_k=\frac{4}{3}{\cal J}_0\beta^2\sin^2\left(\gamma+\frac{2}{3}\pi k\right),
  \end{equation}
are in accordance ($\beta$ and $\gamma$ are the standard deformation parameters in Lund convention \cite{Bo75}).
Fig. 1 shows the ratios given by  two expressions as functions of the triaxiality parameter. Also shown are the ratios calculated 
by means of the microscopic cranking model \cite{TAC} based on the modified oscillator potential. Calculations based on the 
Woods-Saxon potential give essentially the same results. The microscopic ratios follow the irrotational ones, where the moment of inertia
of the short axes is systematically larger. The deviation increases with reduction of the pair correlations. It is to be underlined that without pairing
 the ratios strongly deviate from the rigid body ones, such that they are in accordance with the fundamental properties of the system.
 The systematic study of the 2$^+_2$ states in even-even nuclei by Allmond and Wood \cite{AW17} provides experimental evidence for 
 a $\gamma$ dependence of the moment of inertia ratios that is close to the one in Fig. 1 (a).

 Frauendorf and D\"onau \cite{FD14} classified the particle-triaxial rotor system as, respectively, transverse or longitudinal 
  when the triaxial potential of the rotor aligns the 
 angular momentum of the particle with a principal axes that is perpendicular to or parallel with the axis with the largest moment of inertia.  Transversality
 or longitudinalitly  are
 reflected by a respective decrease or increase of the excitation energy of the wobbling band with the total angular momentum of the system.  
 Accordingly, $^{163}$Lu and $^{135}$Pr are transverse, because the odd proton's angular momentum tends to be aligned with the short axis while the 
 medium axis has the maximal moment of inertia. Using the order ${\cal J}_m>{\cal J}_s>{\cal J}_l$ found by the microscopic calculations, Frauendorf and D\"onau
  were able to account for the observed energies and transition probabilities.  
   
  For their version of the particle-triaxial-rotor model \cite{Tanabe10,Tanabe17}, Tanabe and Sugawara-Tanabe assume the rigid body ratios  (1), 
  which assign the largest  moment of inertia to the short axis. This scenario  (longitudinal   according to \cite{FD14})  results
  in an increase of the wobbling frequency with angular momentum. An angular momentum dependent scaling factor is multiplied to 
  all three moments of inertia, such that the experimentally observed decrease of the wobbling frequency is achieved.   
 Adjusting the triaxiality parameter $\gamma$ the authors  are able to fairly well describe the experimental information on transition probabilities.  
 However,  the striking contradiction with the preceding discussion of the ratios between the three moments of inertia raises serious concerns about the
 suggested scenario. As seen in Fig. 1 (c),  the three rigid body moments of inertia are almost the same for the core of $^{135}$Pr. 
 The $\gamma$ dependence of the rigid-body moments of inertia is obviously wrong for weakly triaxial  and axial nuclei.   
 
 In Ref. \cite{Tanabe17}, Tanabe and Sugawara-Tanabe use a small amplitude approximation to the full particle rotor system
 to study the stability of transverse wobbling for irrotationlal flow ratios between the moments of inertia.
 They conclude with: " There is no wobbling mode around the axis with medium MoI in the particle-rotor model even with the hydrodynamical
MoI..." (that is no transverse wobbling). Such a general conclusion is incorrect. The authors considered only weakly deformed nuclei as $^{135}$Pr, for which 
 they find instability of transverse wobbling  for $I>13/2$. Frauendorf and D\"onau demonstrated in Ref. \cite{FD14} 
 that for   the exact particle rotor solution transverse wobbling becomes only unstable  for $I>17/2$.
 When the moment of inertia of the short axis  is increased to ${\cal J}_s=0.6{\cal J}_m$ the instability moves up to $I=29/2$ where it is observed in experiment.   
 The microscope calculations 
  indicate a larger moment of inertia of the short axis than irrotational flow. The full particle rotor calculations  
 give stable transverse wobbling for the strongly deformed  nucleus $^{163}$Lu for both irrotational flow and microscopic moments of inertia.
 
 To summarize, stable transverse wobbling does exist and the assumption of rigid body ratios between the moments of inertia  of the triaxial rotor core in Refs. \cite{Tanabe10,Tanabe17}
 contradicts basic concepts of  quantal rotation.

\newpage
\onecolumngrid
\begin{figure}[h]
\begin{center}
\includegraphics[width=1.8\linewidth] {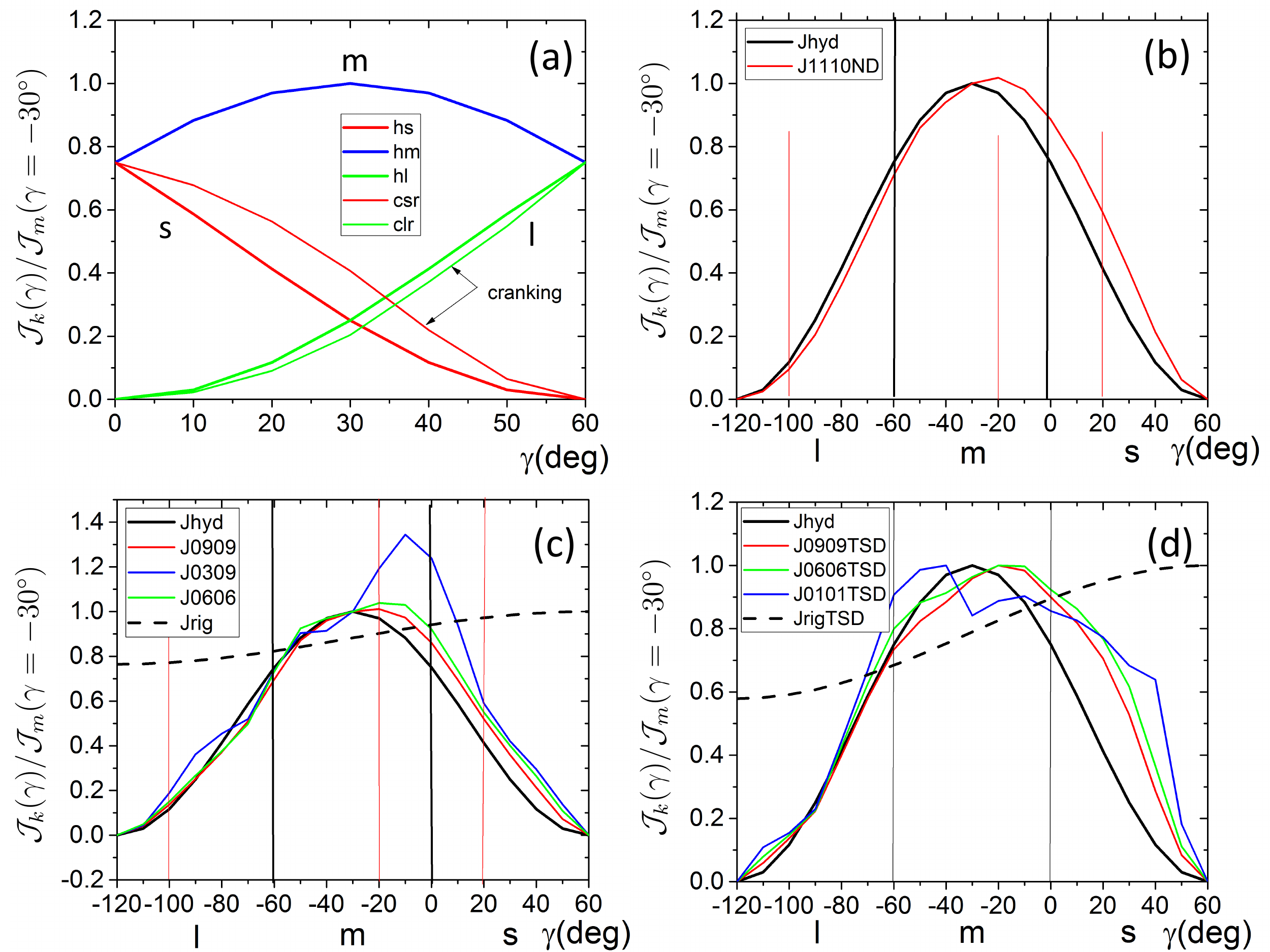} 
\end{center}
 \end{figure}  
 \onecolumngrid
\begin{figure}[h]
\vspace*{-1.3cm}
 \caption{  The moments of inertia of the three principal axes as function of the triaxiality parameter $\gamma$. Thick black curves: irrotational flow values Eq. (2);
 thick dashed curves: rigid body values Eq. (1); thin colored curves: microscopic values obtained by cranking calculations. In panel (a) {\bf hs, hm hl} denote the 
 irrotational flow values and {\bf csr, clr} the cranking values, which are scaled by the factor ${\cal J}^{hyd}_m(\gamma)/{\cal J}^{crank}_m(\gamma)$.  
 The numbers in the legends quote the pairing
 strength. J1110 means $\Delta_p=1.1$MeV, $\Delta_n=1.0$MeV; J0309 means $\Delta_p=0.3$MeV, $\Delta_n=0.6$MeV; etc. 
 Panels (a) and (b) show the same calculations
 for $Z=68$, $N=96$, $\varepsilon=0.25$; panel (c) for $Z=38$, $N=76$, $\varepsilon=0.2$; and panel (d) for $Z=68$, $N=96$, $\varepsilon=0.4$.
 In panels (b-d) the three moments of inertia  are shown in the three $\gamma$ intervals: long $-120^\circ \leq \gamma \leq -60^\circ$, with $\gamma \rightarrow \gamma+120^\circ$;
 medium  $-60^\circ \leq \gamma \leq 0^\circ$, with $\gamma \rightarrow -\gamma$; short   $0^\circ \leq \gamma \leq 60^\circ$; compare (a) with (b). The intersections 
 of the red vertical lines indicate the moments of inertia of the three axes for $\gamma=20^\circ$.}
 \end{figure}
 \twocolumngrid

\end{document}